\documentclass[twocolumn,nofootinbib,showpacs,showkeys,preprintnumbers,superscriptaddress,aps] {revtex4}

\usepackage{amssymb,amsmath,amsfonts,amsbsy,graphicx}
\usepackage{color}
\usepackage{enumerate} 
\usepackage{ulem}

\newcommand{\dis}[1]{\begin{equation}\begin{split}#1\end{split}\end{equation}}
\newcommand{\be}{\begin{equation}}
\newcommand{\ee}{\end{equation}}

\newcommand{\eq}[1]{Eq.~(\ref{#1})}

\newcommand{\bfrac}[2]{{\left(\frac{#1}{#2} \right)  }}\newcommand{\VEV}[1]{\langle #1 \rangle}
\newcommand{\Mp}{M_P}
\newcommand{\tev}{\,\textrm{TeV}}
\newcommand{\gev}{\,\textrm{GeV}}

\newcommand{\ev}{\,\textrm{eV}}

\newcommand{\tildenu}{{\tilde{\nu}}}
\newcommand{\tildeN}{{\tilde{N}}}

\begin{document}

\title{Thermal production of light Dirac right-handed sneutrino dark matter }

\author{Ki-Young Choi}
\email{kiyoungchoi@skku.edu}
\affiliation{Department of Physics, BK21 Physics Research Division, Institute of Basic Science, Sungkyunkwan University, Suwon 440-746, South Korea}

\author{Jongkuk Kim}
\email{jongkukkim@skku.edu}
\affiliation{Department of Physics, BK21 Physics Research Division, Institute of Basic Science, Sungkyunkwan University, Suwon 440-746, South Korea}
 
\author{Osamu Seto}
 \email{seto@particle.sci.hokudai.ac.jp}
 \affiliation{Institute for the Advancement of Higher Education, Hokkaido University, Sapporo 060-0817, Japan}
 \affiliation{Department of Physics, Hokkaido University, Sapporo 060-0810, Japan}

\begin{abstract}
We consider the production of right-handed (RH) sneutrino dark matter in a model of Dirac neutrino where neutrino Yukawa coupling constants are very small. Dark matter RH sneutrinos are produced by scatterings and decays of thermal particles in the early Universe without reaching thermal equilibrium due to the small Yukawa couplings. We show that not only decays of thermal particles but also the thermal scatterings can be a dominant source as well as non-thermal production in a scenario with light sneutrinos and charged sleptons while other supersymmetric particles are heavy. We also discuss the cosmological implications of this scenario.
\end{abstract}

\pacs{ }

\preprint{EPHOU-18-010}

\maketitle

\section{Introduction}                                           
\label{introduction}     

Many experiments have established that neutrinos have non-vanishing masses, while the standard model (SM) was constructed assuming massless neutrinos. One of the simplest extension of the standard model for having massive neutrinos is to introduce the right-handed (RH) neutrinos and their Yukawa couplings.
In the see-saw mechanism~\cite{Seesaw1,Seesaw2,Seesaw3}, heavy Majorana masses ($\gg 100 \gev$) of RH neutrinos are introduced and the light neutrino masses ($\lesssim \ev$) are obtained by the mass hierarchy.
However, even without the see-saw mechanism, the light neutrino mass can be obtained from the Dirac mass, as in the same way of quarks and leptons, if the neutrino Yukawa coupling constants are very small of $\mathcal{O}(10^{-13}-10^{-12})$. 
After the electroweak symmetry breaking, the small Yukawa couplings give Dirac masses to the neutrinos.

In the theory of supersymmetry (SUSY), there exists the superpartner of the RH neutrinos, RH sneutrinos ($\tilde{N}$). Their masses come from the soft SUSY breaking and in many cases are of the order of the gravitino mass. Therefore, it is possible that the  RH sneutrino is the lightest SUSY particle (LSP), stable due to the R-parity conservation and a good candidate for dark matter.

The possibility of the RH sneutrino as dark matter in the pure Dirac type neutrino and those productions through decays of various thermalized superparticles was pointed out by Asaka et al~\cite{Asaka:2005cn}. 
RH sneutrinos are not thermalized in the early Universe, because the interactions of RH sneutrinos are extremely weak.
However, it has been shown that the right amount of dark matter can be produced in the parameter region where left-handed (LH) and RH sneutrinos are degenerate and sleptons are fairly light~\cite{Asaka:2005cn}.
The dark matter abundance estimation was extended by including the non-thermal production of RH sneutrinos from decays of the next-to LSP (NLSP) after its freeze-out~\cite{Asaka:2006fs,Page:2007sh}.
Such a long-lived NLSP scenario is constrained by the big bang nucleosynthesis (BBN)~\cite{Ishiwata:2009gs}. 
Possible collider signatures for the stau NLSP have been investigated~\cite{Banerjee:2016uyt,Banerjee:2018uut}.
For the long-lived stau NLSP case, the lower mass bound for the stau (slepton) is reported as
 $\gtrsim 290$ $(380)$ GeV by the ATLAS~\cite{ATLAS:2014fka} and $\gtrsim 340$ GeV by the CMS~\cite{Chatrchyan:2013oca,Khachatryan:2016sfv}.
 However this is not applicable to our scenario which we will describe below.

In this article, we consider the production of RH sneutrinos by scatterings in the thermal plasma.
Usually and in previous studies~\cite{Asaka:2005cn,Asaka:2006fs,Page:2007sh}, it has been considered that the scattering contribution is subdominant compared to the thermal production from decays or non-thermal production by decay of freeze out particles. However, we find that it can be comparable or even a dominant source, because there are huge number of scattering modes for the production of RH sneutrinos. Especially, such a dominant thermal production from scatterings realizes under the mass spectrum where the charged sleptons and sneutrinos are light and other SUSY particles are heavy, which is indicated by the observation of the SM-like Higgs boson with the mass of $125$ GeV~\cite{Aad:2012tfa,Chatrchyan:2012xdj}. 
In our analysis, for the illustration, we adopt the mass spectrum that one of the RH sneutrinos is the LSP, whose mass
 is close to those of the LH sneutrinos, LH charged sleptons and as light as around $\mathcal{O}(10^2)~\gev$ and other SUSY particles are as heavy as about $1 - 5$ \tev  except one light Higgsino-like neutralino around 700 GeV.

We also consider the cosmological issues in this scenario. Since the LH sneutrinos are slightly heavier than RH sneutrinos, LH sneutrinos may decay relatively late around or after the BBN,
in which case strong constraint can be imposed on the model. 
Heavier RH sneutrinos may decay in the present Universe to the lightest RH sneutrinos with possible observational signals in the cosmic ray observation.

\section{Model}
\label{model}

We consider the extension of the minimal supersymmetric standard model (MSSM) by adding three singlet superfields, $\nu_R^c$, for RH neutrinos with a superpotential in the lepton and Higgs sector;
\dis{
W \ni y_e L\cdot H_d \, E_R^c + y_\nu L\cdot H_u \, \nu_R^c +\mu H_u\cdot H_d  + \textrm{H.c.}, \label{superW}
}
where $H_u=(H_u^+,H_u^0)$,  $H_d=(H_d^0,H_d^-)$ are up-type and down-type Higgs doublet, $L=(\nu_L,l_L)$ is a LH lepton doublet,    $E_R^c$ is a RH charged lepton and $\mu$ in the third term is the SUSY invariant Higgsino mass. Here, we omitted the generation indices.

After the electroweak symmetry breaking, the neutrinos acquire masses with the small Yukawa couplings as
\dis{
y_\nu \sin\beta  \simeq 3.0 \times 10^{-13} \bfrac{m_\nu^2}{2.8\times 10^{-3} \, \ev^2}^{1/2},
}
 with the vacuum expectation value (VEV) of Higgs field $\VEV{H_u} = v_0/\sqrt{2} \sin\beta$, $v_0\simeq 246\gev$ and $\tan\beta=\VEV{H_u} /\VEV{H_d} $. 
In our evaluation of the RH sneutrino production, we take the temperature dependence of Higgs VEV into account, following Refs.~\cite{Asaka:2006fs}, as 
\dis{
v=v(T) = v_0\sqrt{1- T^2/T_c^2 }  \qquad \mathrm{for} \quad T<T_c,
}
with $T_c=147\gev$ according to the SM-like Higgs boson mass $m_h\simeq 126\gev$.
For $T>T_c$, the Higgs VEV vanishes.

The F-term potential includes 
\dis{
V_F\ni \left| y_\nu \tilde{\nu}_L \tilde{\nu}_R^* -\mu H_d^0 \right|^2 +  \left| y_\nu \tilde{l}_L \tilde{\nu}_R^* +\mu H_d^- \right|^2. \label{V_F}
}
Soft SUSY breaking terms for leptons and Higgs fields are given by
\begin{align}
V_{\rm soft} =& m_{H_u}^2|H_u|^2 + m_{H_d}^2|H_d|^2
    + m_{\tilde{L}}^2|\tilde{L}|^2+ m_{\tilde{\nu}_R}^2|\tilde{\nu}_R|^2   \nonumber \\
 & + ( y_l A_l \tilde{L} \cdot H_d \tilde{E}_R +y_{\nu} A_{\nu} \tilde{L} \cdot H_{u} \tilde{\nu}_R^*  - B \mu H_u \cdot H_d \nonumber \\
 &  + \textrm{H.c.} ). \label{Vsoft}
\end{align}
The soft trilinear coupling $A_\nu$-term in~\eq{Vsoft} and the first term in~\eq{V_F} together generate mixing between LH and RH sneutrinos. 
The mixing angle ${\Theta}$ is given by
\dis{
\tan2\Theta(T) \simeq \frac{2m_\nu(T)|\cot\beta \mu-  A_\nu^*|}{ m^2_{\tilde{\nu}}(T)- m_{\tilde{N}}^2}.
\label{mixing}
}
Here, $\tildenu$ and $\tildeN$ denote the mass eigenstates of almost LH and RH sneutrinos, respectively.
Due to small mixing of sneutrinos, in the most cases except e.g., the production in the early Universe as we will show,
 the LH and RH sneutrinos themselves can be practically regarded as their mass eigenstates with eigenvalues,
\dis{
m^2_{\tilde{\nu}} \simeq m_{\tilde{L}}^2 + \frac{m^2_Z}{2} \cos2\beta,  \qquad m_{\tilde{N}}^2  \simeq m_{\tilde{\nu}_R}^2,
}
where the second term in $m^2_{\tilde{\nu}}$ comes from the $D$-term contribution.
In the followings, we use a dimensionless parameter $a_\nu = A_\nu/m_{\tilde{L}}$ to parameterize $A_\nu$. 

The relevant interactions for the production of RH sneutrinos are the Yukawa interactions in \eq{superW}, the corresponding trilinear couplings in \eq{Vsoft}, cross terms involving the $\mu$-term in \eq{V_F}, and the resultant mixing between LH and RH in \eq{mixing}.
Even though these interactions are very weak, they are important and must be kept for the production of RH sneutrinos, since they are the dominant interactions of RH sneutrinos with the SM sector. 
In the case of degenerate masses between LH and RH sneutrinos, the mixing can be the dominant source among them.
In Tab.~\ref{table:couplings}, we summarize the interactions of RH sneutrnios with other MSSM particles.

\begin{table}
\begin{center}
\begin{tabular}{|c|c|} \hline
 \makebox[25mm][c]{Interaction} &  \makebox[30mm][c]{Coupling}  \hfill \\
\hline \hline
$\overline{\chi}_a^0  \nu\tilde{N}^*$ & $\left[- y_\nu N_{a3} + \frac{ e\sin\Theta}{\sqrt{2}s_wc_w} (s_w N_{a1} - c_w N_{a2}) \right] P_L$\\
\hline
$\overline{\chi}^-_a   l^- \tilde{N}^*$ & $\left[ y_\nu Z_{a2}  - \frac{e}{s_w}Z_{a1} \sin\Theta \right] P_L$\\
\hline
$\tilde{\nu} \tilde{N}^* Z_\mu$&$\frac{-ie}{2s_wc_w}(p-k)^\mu \sin\Theta$\\
\hline
$\tilde{l}^- \tilde{N}^* W_\mu^+$&$-i\frac{e}{\sqrt{2}s_w}(p-k)^\mu \sin\Theta$\\
\hline
$\tilde{\nu} h^0 \tilde{N}^*$&$-y_\nu \frac{1}{\sqrt2}(A_\nu \cos\alpha +\mu\sin\alpha )$\\
\hline
$\tilde{\nu} H^0 \tilde{N}^*$&$-y_\nu \frac{1}{\sqrt2}(A_\nu \sin\alpha -\mu\cos\alpha )$\\
\hline
$\tilde{\nu} A^0 \tilde{N}^*$&$iy_\nu \frac{1}{\sqrt2}(A_\nu \sin\beta -\mu\cos\beta )$\\
\hline
$\tilde{l}^- H^+ \tilde{N}^*$ & $y_\nu (A_\nu \cos\beta+\mu\sin\beta)$\\
\hline
\end{tabular}\caption{The interactions of RH sneutrinos. Here $s_w=\sin\theta_W$ and $c_w=\cos\theta_W$ where $\theta_W$ is the Weinberg mixing angle.  
The neutralino mass eigenstates are given by a linear combination of the bino $\tilde{B}$, the neutral wino $\tilde{W}_3^0$ and two neutral Higgsinos $\tilde{H}_u^0$ and $\tilde{H}_d^0$, $\tilde{\chi}_a = N_{a1}\tilde{B} + N_{a2}\tilde{W_3^0} + N_{a3}\tilde{H}_u^0+ N_{a4} \tilde{H}_d^0 $, where $N_{ab}$ is the mixing matrix for neutralino mass. $Z_{ab}$ is the mixing matrix for chargino mass eigenstates, $\tilde{\chi}_a^-=Z_{a1} \tilde{W}^- + Z_{a2}\tilde{H}_u^-$.  The light and heavy Higgs mass eigenstates are $h^0= -\sin\alpha \sqrt2 \textrm{Re}(H_d^0) + \cos\alpha \sqrt2 \textrm{Re}(H_u^0) $ and $H^0= \cos\alpha \sqrt2 \textrm{Re}(H_d^0) + \sin\alpha \sqrt2 \textrm{Re}(H_u^0)$ with $\tan2\alpha = \tan 2\beta \bfrac{m_A^2+M_Z^2}{m_A^2-M_Z^2}$, and $H^+ = \sin \beta (H_d^-)^* + \cos\beta H_u^+$ and  $A^0=\cos\beta \sqrt2\textrm{Im} (H_d^0) - \sin\beta \sqrt2\textrm{Im} (H_u^0)$. 
}
\label{table:couplings}
\end{center}
\end{table}

\section{Thermal and non-thermal production}
\label{production}

Cosmic microwave background (CMB) anisotropy and the large scale structure formation requires that the relic density of dark matter at present is~\cite{Ade:2015xua}
\dis{
\Omega_{\rm DM} h^2 = 0.1188\pm 0.0010.
}
Dark matter should have been produced in the early Universe.
The early Universe was dominated by hot thermal particles that maintains thermal equilibrium through interactions in the expanding Universe.
However, the very weakly interacting particles cannot be in equilibrium in a given Hubble time.
The coupling of RH sneutrinos to the SM particles are very small, which is suppressed by the Yukawa coupling or by the mixing between LH and RH sneutrinos. Their interactions are not enough to make the RH sneutrinos in the thermal equilibrium. Instead, only small amount of RH sneutrinos are produced at low temperature during freeze-in from the thermal particles by scatterings or decays, which is the property of feebly interacting massive particles (FIMP).  

For heavy RH sneutrinos with GeV mass, the small amount of RH sneutrinos of the number density around $10^{-9}$ times of photon number density is enough to explain dark matter.
The present relic density of non-relativistic RH sneutrinos dark matter with the mass $m_{\tildeN}$ can be expressed by
\dis{
\Omega_{\tildeN} h^2 \simeq 0.28 \bfrac{Y_{\tildeN}}{10^{-11}}\bfrac{m_{\tildeN}}{100\gev},
}
where the abundance yield $Y$ is defined as $Y_{\tildeN} \equiv n_\tildeN/s$ with the number density $n_\tildeN$, the entropy density $s=(2\pi^2/45)g_{s*} T^3$ and the relativistic degrees of freedom in the entropy density $g_{s*}$.

There are two mechanisms for producing RH sneutrinos:
{\it thermal production} (TP) and {\it non-thermal production} (NTP). TP includes the production via scatterings or decay processes of the particles in the thermal equilibrium. NTP is the production from particles that are already out of the thermal bath, typically by the late time decay of NLSPs after their freeze out.

\begin{table}
\begin{center}
\begin{tabular}{|c|c|} \hline
Scattering Process&Number of modes \\
\hline\hline
$ Z + \tilde{\ell}_j \to W^- +  \tilde{N}_j  $ & $3\times 3_g$ \\
\hline
$W^\pm + \tilde{\nu}_j \to W^\pm + \tilde{N}_j $ &$2\times3_g\times2_{\pm}$ \\
\hline
$u_i (\nu_i )+ \tilde{\ell}_j \to d_k (\ell_i )+ \tilde{N}_j$  & $3\times3_g^2 (3\times3_g^2\times3_{susy})$\\
\hline
$q_i (\ell_i ,\nu_i)+ \tilde{\nu}_j \to q_i (\ell_i ,\nu_i)+ \tilde{N}_j$  & $3\times3_g^2 (3\times3_g^2\times3_{susy}\times 2_{l,\nu})$ \\
\hline
$W^+ + \tilde{\ell}_j \to h (\gamma) + \tilde{N}_j $&$3\times3_g \times 2_{h,\gamma}$ \\
\hline
$Z + \tilde{\nu}_j \to h (\gamma) + \tilde{N}_j $ &$3\times3_g \times 2_{h,\gamma}$  \\

\hline
\end{tabular}\caption{Relevant scattering processes for thermal production of $\tilde{N}$ with light external particles. $i,j,k$ denote the generation. In the second column, we show the number of modes for the process in the left column. There are additional processes of charge conjugate.  } 
\label{table:scatterings}
\end{center}
\end{table}

\subsection{Thermal production}
The number density of RH sneutrinos from TP can be obtained by solving the Boltzmann equation with scattering and decay processes. Since the number density of the RH sneutrinos is well below the equilibrium value initially, which is well justified after inflation, we can ignore the inverse processes. Thus, we have
\dis{
\frac{dn_{\tilde{N}}}{dt}+3H n_{\tilde{N}} \simeq & \sum_{i,j} \VEV{\sigma (i+j\rightarrow \tilde{N} + \cdots)v_{rel}}n_in _j\\
&+ \sum_i \VEV{\Gamma(i \rightarrow \tilde{N} + \cdots )}n_i , \label{Boltz}
}
where $H$ is the Hubble parameter and $n_i$ is the comoving number density of $i$-th particle. The first and the second term in the RHS are due to the two-body scatterings and the decay of $i$-th particle into $\tilde{N}$ and other species. 
Here '$\VEV{\cdots}$' denotes the thermal average, $\sigma v_{rel}$ is the product of the scattering cross section and the relative velocity, and $\Gamma$ is the decay rate for a process presented in round brackets.

\begin{figure*}[!t]
\begin{center}
\begin{tabular}{cc} 
 \includegraphics[width=0.4\textwidth]{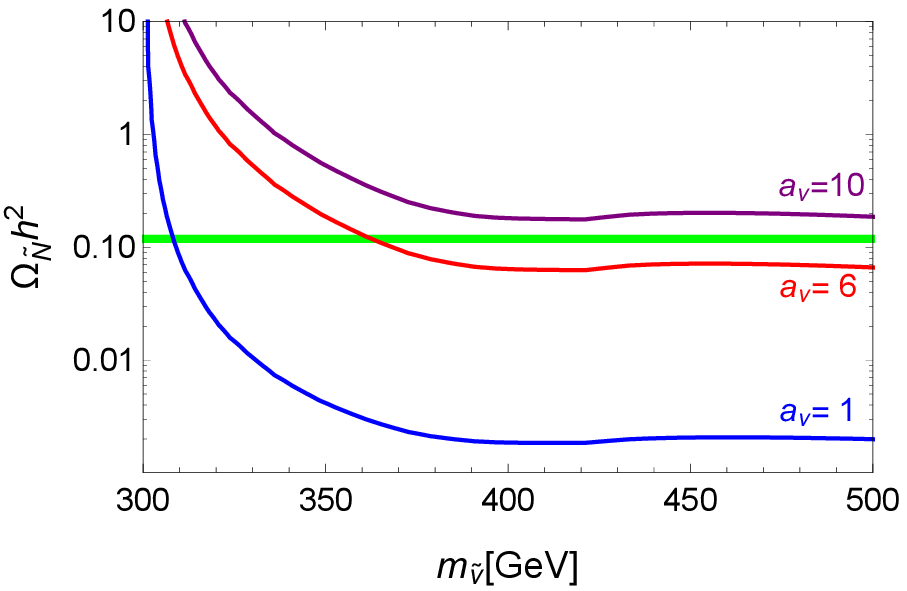}
 &
 \includegraphics[width=0.4\textwidth]{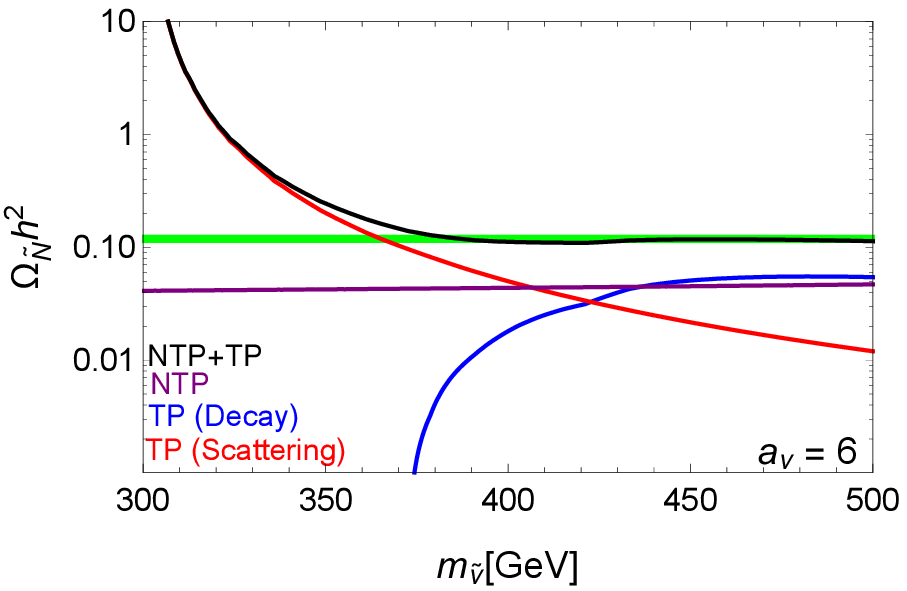}
   \end{tabular}
\end{center}
\caption{The abundance of RH sneutrinos from thermal production (scattering + decay) with $m_{\tildeN}=300$ GeV (Left) and the total abundance including nonthermal production with breakdown of contribution for $a_\nu=6$ (Right). Three lines in the left panel correspond to different values of $a_\nu=1,6,10$ from bottom to top. } 
\label{fig:relicdensity}
\end{figure*}

The solution for the Boltzmann equation~\eq{Boltz} can be expressed in terms of the abundance of RH sneutrinos,
 which has two contributions from scattering and decay respectively, as
\dis{
Y_\tildeN^\mathrm{TP} =  \sum_{ij} Y_{ij}^\mathrm{scat} + \sum_i Y_i^\mathrm{dec},
}
 with
\dis{
Y_{ij}^\mathrm{scat} &=  \int_{T_0}^{T_R} dT \frac{\VEV{\sigma (i+j\rightarrow \tilde{N} + \cdots)v_{rel}}n_in _j}{sHT},\\ 
Y_i^\mathrm{dec} &=   \int_{T_0}^{T_R} dT \frac{\VEV{\Gamma(i \rightarrow \tilde{N} + \cdots )}n_i}{sHT}.
}
Here $T_0$ is the present temperature and $T_R$ is the reheating temperature after inflation.
Especially for two-body scatterings, the abundance of dark matter from scattering is given by~\cite{Choi:1999xm}
\dis{
Y^{\rm scat} &=\int_{T_0}^{T_R} \frac{\VEV{\sigma v } n_in_j }{s H T} dT \\
&= \frac{\bar{g} \Mp}{16\pi^4} \int_{t_R}^\infty dt \, t^3 K_1(t) \int_{(m_1+m_2)}^{t T_R} d(\sqrt{s}) \sigma(s) f(s), \label{Yscatt}
}
with $\bar{g}=135\sqrt{10}/(2\pi^3g_*^{3/2})$, $t_R=(m_1+m_2)/T_R$ and
\dis{
f(s)=\left[ \frac{(s-m_1^2 -m_2^2)^2 -4 m_1^2 m_2^2}{s^2} \right],
}
where $m_1$ and $m_2$ are the masses of initial state particles, and $T_0=0$ is taken as usual.

The scattering processes relevant in our scenario are listed in Tab.~\ref{table:scatterings}.
Here, we show processes dominant for the production of RH sneutrinos, which include  the light initial particles such as LH and RH sneutrinos, the charged sleptons, 
 as well as the number of scattering modes possible for each process in the right column. 
The number of modes include the crossing of the process, three generations, and supersymmetric counterparts. We note that there is another factor 2 from the processes of charge conjugate.

The thermal production from decays, when the temperature of the early Universe was much higher than the decaying particles, can be well approximated by~\cite{Choi:1999xm}
\dis{
Y^\mathrm{decay}\simeq \sum_i \frac{3\zeta(5)\bar{g} \Mp g_i \Gamma_i}{4\pi m_i^2}, \label{Ydecay}
}
with $g_i$ being the degrees of freedom in the rest frame and $\bar{g}=135\sqrt{10}/(2\pi^3g_*^{3/2})$.
The decay modes which produce RH sneutrinos are well summarized in Refs.~\cite{Asaka:2005cn,Asaka:2006fs}.
Note that the abundance from decay, $Y^\mathrm{decay}$, is proportional to $\Gamma_i/m_i^2$ in \eq{Ydecay}.
Therefore, for the decay whose rate is given as $\Gamma_i \propto m_i$, the resultant abundance by those is suppressed for such heavy particles. This is the case for neutralinos and charginos in our scenario. 

Since we consider the temperature dependence of Higgs VEV, the scattering or decay through mixing are suppressed
at high temperature much larger than $v_c$.

\subsection{Non-thermal production and relic density}

The NTP of RH sneutrino is obtained from the decay of LH sneutrino NLSPs.
The number of produced RH sneutrino is the same as that of the original NLSP and thus we get
\dis{
Y_{\tildeN}^\mathrm{NTP} = Y_{NLSP}, 
}
where $Y_{NLSP}=n_{NLSP}/s$ is the abundance of the LH sneutrinos. 

In our scenario, the two-body decay rate for $\tilde{\nu}\rightarrow\tilde{N} \, Z $ is given by
\dis{
\Gamma_{\tilde{\nu}}^{2-body} &\simeq \Gamma(\tilde{\nu}\rightarrow\tilde{N} \, Z) \\
 &=  \frac{\sin^2\Theta}{32\pi} \frac{A_\nu^2}{m_{\tildenu}} \beta_f^3,
}
with a kinematic factor
\dis{
\beta_f^2 = \frac{1}{m_x^4}[m_x^4 - 2(m_\tildeN^2+m_y^2)m_x^2 + (m_{\tildeN}^2 - m_y^2)^2],
} 
for the processes $x\rightarrow \tildeN y$.
If the two-body decay is kinematically not allowed, the three-body decay mediated by $Z$ boson with the decay rate
\dis{
\Gamma_{\tilde{\nu}}^{3-body} &\simeq \sum_f  \Gamma(\tilde{\nu} \rightarrow \tilde{N} + \bar{f} +f )\\
&  \simeq 0.3 \times \frac{\alpha_{\rm em}^2 \sin^2\Theta \,m_{\tilde{\nu}}^5}{\pi m_Z^4} g\left(\frac{m_{\tildeN}}{m_\tildenu}\right),
} 
dominates.
Here, $\alpha_{\rm em}$ is the fine structure constant,
\dis{
g(x) = (1-x^4)(1-8x^2+x^4) -24x^4\log x.
}
is an auxiliary function, and $f$ denotes all kinematically allowed leptons and quarks.
For $1-x\ll $1, the limit gives $g(x)\simeq 64(1-x)^5/5$.
For the degenerate case, the lifetime of a LH sneutrino from three-body decay is estimated as
\dis{
\tau _{\tilde{\nu}}  \simeq  10^3\sec \bfrac{300\gev}{m_{\tilde{N}}}\bfrac{0.1\ev}{m_\nu}^2\bfrac{0.1}{\delta}^3\bfrac{1\tev}{|\cot\beta\mu-A^*|}^2,
\label{tau_L}
}
where $\delta = (m_{\tilde{\nu}}-m_\tildeN)/m_\tildeN \ll 1$ is assumed.
Due to the small Yukawa couplings and degenerate masses, LH sneutrinos would decay late in the Universe after BBN, which may disturb the abundance of the light elements or affect the CMB anisotropy. 
We will examine this issue later.

The total relic abundance of RH sneutrinos is then the sum of TP and NTP of both $\tildeN$ and its antiparticles, $\tildeN^*$,
\dis{
Y_{\tildeN+\tildeN^*} = 2(Y^{TP}_{\tildeN} + Y^{NTP}_{\tildeN}).
}
In comparison with the critical density, the relic density of RH sneutrinos is then given by
\dis{
\Omega_{\tildeN+\tildeN^*} = 2.8\times 10^{10} \bfrac{m_\tildeN}{100\gev}Y_{\tildeN+\tildeN^*}.
}
Then the abundance of  DM  is the sum of three generations of right-handed sneutrinos.\dis{
\Omega_{DM}=\sum_i\Omega_{\tildeN_i+\tildeN^*_i}.
}
Note that the heavier RH sneutrinos may decay in the present Universe and give signatures in the indirect detection experiments~\cite{Demir:2009kc}.

\begin{figure*}[!t]
\begin{center}
\begin{tabular}{cc} 
 \includegraphics[width=0.45\textwidth]{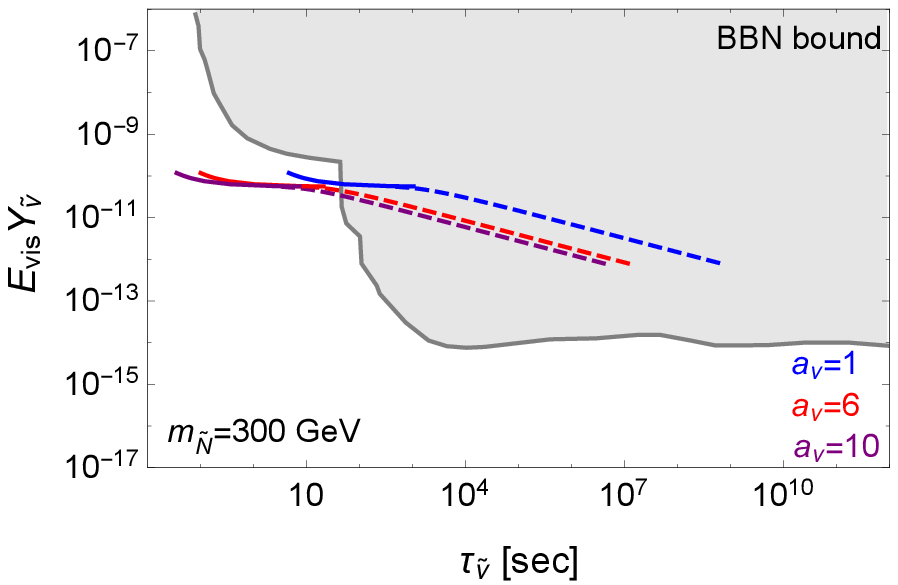}  
 &
 \includegraphics[width=0.45\textwidth]{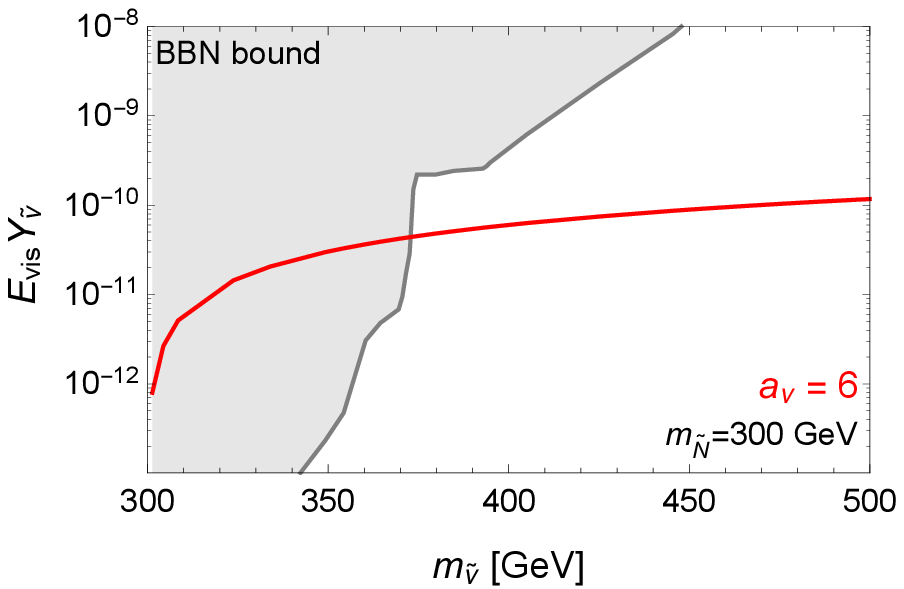}  
 \end{tabular}
\end{center}
\caption{Left: The BBN constraint on the lifetime vs the yield plane for $m_{\tilde{N}}=300$ GeV. 
The BBN disallowed region has gray shading. 
The lines represent the relation between $E_{vis}Y_{\tilde{\nu}}$ which is determined by
 the NLSP freeze out and the lifetime for different $a_{\nu}$. 
In this plot, we vary $m_{\tilde{\nu}}$ for a range $301.3$ GeV $ \leq m_{\tilde{\nu}} \leq 500 $ GeV.
The solid and dashed lines stand for the two-body and three-body decay, respectively.  Right: The same as left panel, but the mass of LH sneutrino is taken as the horizontal axis.}
\label{BBN}
\end{figure*}

\section{Production of right handed sneutrinos in the early Universe}
\label{relicdensity}

We consider that the LH sleptons and RH sneutrinos are light and degenerate with mass around $100$ GeV and all other SUSY particles are heavy between $1$ TeV and $5$ TeV to be consistent with LHC bounds. We note, for information, the typical mass spectrum we use are as follows; the mass of the lightest neutralino is Higgsino-like with the mass about $700$ GeV, those of squarks are about $2$ TeV to be consistent with $m_h \simeq 126$ GeV and the gluino mass is about $5$ TeV. 
In this case, the thermal production from decay can be sub-dominant and the production from scatterings become dominant sources for producing RH sneutrinos.

In the previous studies, the scattering processes have been neglected since they are considered to be subdominant and cannot produce enough RH snuetrinos. It is true when we consider only single scattering mode. However as we can see in Tab.~\ref{table:scatterings}  we find that there are more than $300$ scattering modes which can contribute to the production
and the total contribution is not negligible anymore. 
In our calculation, the abundance is computed by using the code \texttt{micrOMEGAs5.0}~\cite{Belanger:2018mqt}.

In Fig.~\ref{fig:relicdensity}, we show the relic density of RH sneutrinos vs LH sneutrino mass for $m_\tildeN=300\gev$.
In the left panel, we show the thermal production from both thermal scatterings and thermal decays for different trilinear couplings $a_\nu=1,6,$ and $10$ for comparison. In the right panel, the total relic density of RH sneutrinos (black line) and contributions from thermal scatterings (red line), thermal decays (blue line) and non-thermal production (purple line) are shown respectively. Here we take $a_\nu=6$.

For a smaller LH sneutrino mass, $m_\tildenu \lesssim 400\gev$, the thermal scatterings dominate the contribution to the relic density of RH sneutrinos.
For $ 400\gev \lesssim m_\tildenu \lesssim 440\gev$, the TP from scatterings become subdominant and the NTP becomes dominant. For $m_\tildenu\gtrsim 440\gev$, TP from decays can be more important, since the 2-body decay of LH sneutrino to RH sneutrino and $Z$-boson, or that of charged slepton to RH sneutrino and $W$-boson opens.

\section{Cosmological and astrophysical implications}
\label{implication}

\subsection{Constraints from BBN}

Due to the small mixing and degeneracy, a LH sneutrino decay to a RH sneutrino after BBN,
 around $1\sec$ to $10^{10} \sec$ as in \eq{tau_L}.
The decay products of LH sneutrinos contains the electromagnetic and/or hadronic particles, and they
can change the abundance of light elements or affect the distribution of CMB.
To avoid these problems, the abundance of LH sneutrinos are constrained depending on its lifetime.
According to Ref.~\cite{Kawasaki:2004yh,Kawasaki:2004qu,Kawasaki:2017bqm},
the product of visible energy produced from decay and the abundance is constrained by
\dis{
E_{vis} Y_{\tilde{\nu}} < 10^{-13} - 10^{-14}~\mathrm{GeV}, 
}
for $\tau= 10^{3} \sec - 10^{10} \sec$.
Considering $E_{vis}\simeq (m_{\tilde{\nu}}^2 - m_\tildeN^2)/2m_{\tilde{\nu}}\simeq 10\gev$, the abundance is constrained by $Y_{\tilde{\nu}}\lesssim 10^{-14} - 10^{-15}$.

In Fig.~\ref{BBN}, we show the constraint from BBN on $E_{vis} Y_{\tildenu}$ for different lifetime (gray region is disallowed) and the prediction in our model for $a_\nu=1,6, 10$ (blue, red, brown lines respectively).
We find that for $a_\nu=1$, it is almost disfavored. For $a_\nu=6$, two-body decay may avoid the BBN constraint. 
For $a_\nu=10$, two-body decay and also small region of three-body decay might be available where the lifetime is short enough as $\tau \lesssim 100 \sec$.
In the right panel, we show the same figure but with the horizontal axis with the mass of LH sneutrino. We can explicitly find that the BBN constraint can be avoided for $m_\tildenu \gtrsim 370\gev$ where 2-body decays open. 

The heavier RH neutrinos may decay at present Universe with the lifetime around $10^{30} \sec$ with charged lepton pairs. This may be detected in the cosmic rays searches~\cite{Asaka:2006fs}.

\subsection{Constraints from collider experiments}

We have considered cases of the mass spectrum with RH sneutrino LSP, LH sneutrino NLSP, and LH charged sleptons next-to-NLSP. 
Masses of those three kinds of particles are about a few hundreds GeV while the most of other particles are heavier than $1$ TeV. Here, we examine the constraints from collider experiments, principally the LHC. With above mass spectrum, we may observe the decay of a LH slepton to a LH sneutrino and $W^{\pm *}$ but not decay of LH sneutrino to RH sneutrino with the long lifetime as we have discussed above. Then, the signal is very soft jet or lepton from $W^{\pm *}$ with a large missing transverse momentum.
LH slepton pair production in our model gives a signal $\ell^+\ell^- +$ missing transverse momentum through $\tilde{\ell}^+\tilde{\ell}^- \rightarrow W^{+ *}W^{- *} \tilde{\nu}\tilde{\nu}^*  \rightarrow  \ell^+\ell^- \nu\bar{\nu} \tilde{\nu}\tilde{\nu}^*$. This is same as one of signals of chargino pair production $\tilde{\chi}_1^+\tilde{\chi}_1^- \rightarrow W^{+ *}W^{- *} \tilde{\chi}_1^0\tilde{\chi}_1^0$ in the MSSM with neutralino LSP. 
The weak bound for charged Higgsino mass as $\gtrsim 145$ GeV has been derived by ATLAS~\cite{Aaboud:2017leg}.
Hence, the mass region $\gtrsim 150$ GeV presented above is so far consistent with the LHC bounds.

\section{Conclusion}                                           
\label{conclusion}     

We have investigated the production of RH sneutrino dark matter in a pure Dirac neutrino model by scatterings and decays in the thermal plasma. 
The main finding is that the scattering processes in thermal bath, which have been overlooked in previous studies, can be the leading processes in  many cases. 

Motivated by null results of SUSY searches and the observed SM-like Higgs boson mass at the LHC, we have considered the mass spectrum that only LH and RH sneutrinos, and LH charged sleptons are as light as $\mathcal{O}(100)$ GeV and other particle masses are a few TeV.
In this case, the production of RH sneutrinos are dominant or comparable to the production by thermal decays or non-thermal production.
For successful BBN, the lifetime of LH sneutrinos need to be smaller than around 100 sec.

\bigskip
\section*{Acknowledgement}
K.-Y.~C. and J.~Kim was supported by the National Research Foundation of Korea(NRF) grant funded by the Korea government(MEST) (NRF-2016R1A2B4012302). K.-Y.~C acknowledge the hospitality at APCTP where part of this work was done. J.~Kim was supported by the National Research Foundation of Korea(NRF) grant funded by the Korea government(MEST) (NRF-2015R1D1A1A01061507,  NRF-2018R1D1A1B07051127). This work was supported under the framework of international cooperation program managed by the National Research Foundation of Korea (NRF-2018K2A9A2A08000127).

\appendix

\end{document}